# SURVEY

## Applications of Multi-Agent Slime Mould Computing

Jeff Jones[*]

*Centre for Unconventional Computing, University of the West of England, Coldharbour Lane, Brisol, BS16 1QY, UK.*

The giant single-celled slime mould *Physarum polycephalum* has inspired rapid developments in unconventional computing substrates since the start of this century. This is primarily due to its simple component parts and the distributed nature of the 'computation' which it approximates during its growth, foraging and adaptation to a changing environment. Slime mould functions as a living embodied computational material which can be influenced (or *programmed*) by the placement of external stimuli. The goal of exploiting this material behaviour for unconventional computation led to the development of a multi-agent approach to the approximation of slime mould behaviour. The basis of the model is a simple dynamical pattern formation mechanism which exhibits self-organised formation and subsequent adaptation of collective transport networks. The system exhibits emergent properties such as relaxation and minimisation and it can be considered as a virtual computing material, influenced by the external application of spatial concentration gradients. In this paper we give an overview of this multi-agent approach to unconventional computing. We describe its computational mechanisms and different generic application domains, together with concrete example applications of material computation. We examine the potential exploitation of the approach for computational geometry, path planning, combinatorial optimisation, data smoothing and statistical applications.

## 1. Slime Mould Computing

The giant amoeboid Myxomycete organism true slime mould *Physarum polycephalum* has proven to be an ideal candidate for research into living unconventional computing substrates. *P. polycephalum* is a giant single-celled organism which can usually be seen with the naked eye (for a comprehensive guide, see [1]). During the plasmodium stage of its complex life cycle it adapts its body plan in response to a range of environmental stimuli (nutrient attractants, repellents, hazards) during its growth, foraging and nutrient consumption. The plasmodium is composed of a transport network of protoplasmic tubes which spontaneously exhibit contractile activity which is harnessed used in the pumping and distribution of nutrients. The organism is remarkable in that its complex behaviour is achieved without any specialised nervous tissue. Control of its behaviour is distributed throughout the simple material comprising the cell and the cell can survive damage, excision or even fusion with another cell.

The plasmodium of slime mould is amorphous in shape and ranges from the microscopic scale to up to many square metres in size. It is a giant single-celled syncytium formed by repeated nuclear division, comprised of a sponge-like actomyosin complex co-occurring in two physical phases. The gel phase is a dense matrix subject to spontaneous contraction and relaxation, under the influence of changing concentrations of intracellular chemicals. The protoplasmic sol phase is transported through the plasmodium by the force generated by the oscillatory contractions within the gel matrix. Protoplasmic flux, and thus the behaviour of the organism, is affected by changes in pressure, temperature, space availability, chemoattractant stimuli and illumination [2], [3], [4], [5], [6], [7], [8]. The *P. polycephalum* plasmodium can thus be regarded as a complex functional material capable of both sensory and motor behaviour. Indeed *P. polycephalum* has been described as a membrane bound reaction-diffusion system in reference to both the complex interactions within the plasmodium and the rich computational potential afforded by its material properties [9].

---

[*]Email: jeff.jones@uwe.ac.uk

Interest in slime mould computing was initiated by the work of Nakagaki et al. who found that the slime mould could approximate the solution to a simple maze problem when the slime mould was inoculated as fragments covering the channels of a patterned maze. These fragments fused over a number of hours and, when the maze start and end points were covered by nutrient oat flakes, the plasmodium spontaneously adapted its transport network. Protoplasmic tubes were removed from redundant (dead end) paths and longer paths, leaving the transport network of the slime mould connected to the start and end points, thus finding a solution to the maze [10].

Subsequent research into the range of computational abilities of slime mould demonstrated that the plasmodium successfully approximates spatial representations of various graph problems. In [11] the authors examined the connectivity of the tube network when the plasmodium was presented with multiple sources of nutrients. They found that the plasmodium constructed networks that combined features of minimum path length (approximating the Steiner tree) and cyclic connectivity (giving resilience to random disconnection of a path). It has since been found that slime mould successfully approximates spatial representations of various graph problems including generation of Voronoi diagrams and collision-free path planning [12], Delaunay triangulation [13], spanning trees [9, 14, 15], proximity graphs [16], convex hulls and concave hulls [17]. These research examples all used the spatial foraging behaviour of the plasmodium to approximate graph problems which are conventionally solved using algorithmic approaches. Methods to control the propagation of the plasmodium using attractants, repellents and light irradiation were investigated by Adamatzky in [18–20].

The oscillatory phenomena and avoidance of light irradiation were exploited by Aono and colleagues for combinatorial optimisation problems [21–23], specifically small instances of the Travelling Salesman Problem, and found that the chaotic behaviour of the internal oscillations helped the plasmodium avoid deadlock situations, preventing the organism from becoming trapped in local minima – behaviour which is useful in terms of computational and biological search strategies. The behaviour of *Physarum* in response to strong long-distance attractant stimuli combined with short-distance repulsive stimuli was found to follow attractor cycles around simple stimuli and limit-cycle motion with more complex stimuli arrangements [24].

It is somewhat traditional in unconventional computing to validate the computational equivalence of a particular computing substrate with the components of classical computing devices [25–27]. It should be stressed that such research is motivated by exploring theoretical computational *potential*, rather than suitability. In [28], the authors demonstrated how a foraging plasmodium of *Physarum* could be used to construct simple logic gates. A similar approach based on the ballistic computing model was implemented using *Physarum* in [29]. The likelihood of ex-tending this approach for more complex adding circuits was explored in simulation in [30] who found that foraging errors were compounded by small delays in signal timing at junctions, rendering the approach infeasible for larger adding circuits. More recently, photoavoidance by *Physarum* was used to implement a range of logic gates [31]. The protoplasmic tubes of the *Physarum* plasmodium have also been shown to act as microfluidic logic gates under mechanical stimulation [32]. The relatively slow growth and propagation of the *Physarum* plasmodium limits its application for logical gates. However, Whiting recently demonstrated an approach whereby logic operations could be approximated by much faster changes in oscillatory streaming frequency [33]. The utilisation of different frequency responses with regard to arena size was used experimentally and in simulation in the proposal to simplify Adder circuits using a quantitative scheme [34].

Slime mould utilises its self-made protoplasmic network to transport nutrients within its cell body. The transport phenomena correspond to transportation networks formed by collectives in other living systems including fungi [35, 36], ants [37] and humans [38]. Since the plasmodial network is a single cell, constructed from 'bottom-up' principles, how does the structure of the plasmodium networks compare to other artificial transport networks which are typically constructed from hierarchical 'top-down' methodologies? The task is somewhat difficult as slime mould is only concerned with survival, rather than solving externally applied problems, however early research into the topic of nature-inspired transport networks using slime mould was performed by Adamatzky and Jones who found that *Physarum* networks closely approximated the major motorway network connecting the most populous UK urban areas [39]. The authors also found that the plasmodium effected an efficient response to simulated disastrous contamination of individual urban areas, implemented by diffusion of salts within the region. The plasmodium migrated away from contaminated regions to relatively unpopulated areas before re-establishing network connectivity when the damaged areas were contamination-free. This study was recently extended to include the major motorway networks in different countries [40], and an intriguing similarity between the historical evolution of human networks (for example, cattle droving trails, iron age trails, Roman roads, modern arterial routes) can be mirrored in the evolution of early stage fine-grained *Physarum* networks to later networks with thicker and more sparse connectivity [41]. The connectivity of *Physarum* networks was also compared with the regional rail system surrounding Tokyo by Tero et al. who, using a novel approach to represent environmental hazards using light irradiation, also found a similar correspondence between the human and plasmodial networks, in terms of distance and connectivity [42].

## 2. Computational Models of Slime Mould

Early models of *Physarum* were focused on individual biological aspects of its behaviour, most notably the generation, coupling, and phase interactions between oscillators within the plasmodium. More recently, the overall behaviour of the organism has been modelled in attempts to discover more about its distributed computation abilities.

Oscillatory phenomena in *Physarum* have been studied using numerical models to represent the interactions between oscillators in one dimensional systems [43–45], typically modelling the contractions generating protoplasmic streaming phenomena and temporal phase interactions arising from the coupling of the chemical oscillators to mechanical material flux. The model used in [43] was used to explain the primitive memory effect observed in *Physarum*, although an alternative explana-tion based upon memristive effects was suggested in [46]. Methods of approximating the alignment of actomyosin fibres within the plasmodium were presented in [47] and, more generally, oscillatory phenomena within cytogels were modelled in [48]. Numerical models have also been utilised in two dimensions to approximate interactions between pattern transitions [49], spatial response to environmental changes [50] and amoeboid movement [51].

The topology of the *Physarum* protoplasmic tube network is influenced by nutrient concentration and distribution, and evolves to achieve a compromise between minimal transport costs and fault tolerance [11]. Since the plasmodium obviously cannot have any global knowledge about the initial or optimal topology, the network must evolve by physical forces acting locally on the protoplasmic transport. Tero et al. have suggested that protoplasmic flux through the network veins may be the physical basis for evolution of the transport network: given flux through two paths, the shorter path will receive more sol flux. By generating an autocatalytic mechanism to reward veins with greater flux (by thickening/widening them) and to apply a cost to veins with less flux (the veins become thinner), shorter veins begin to predominate as the network evolves. This approach was used for the mathematical model of *Physarum* network behaviour to solve path planning problems [52, 53]. This method indirectly supports the reaction-diffusion inspired notions of local activation (enhancement of shorter tube paths) and lateral inhibition (weakening of longer tube paths). The starting point for the model of Tero et al. is a randomly connected protoplasmic tube network, surrounding a number of nutrient sources (network nodes) which act as sources and sinks of flux. By beginning with a complete network the Tero model, although successful in generating impressive solutions to network problems, sidesteps the issues and mechanisms of initial network formation, plasmodium growth, foraging, and adaptation to a changing nutrient environment.

The flux model of Tero et al. has been adapted to generate solutions to the Steiner tree problem. In the first stage, the tube network is minimised using the flux model initialised with a random initial network. This is then followed by a second stage which attempts to optimise the network structure. Because the original node positions are already known, the second stage simply removes meandering paths connecting the original nodes and the Steiner points, replacing them with straight connections [54]. The flux model was also used in a study comparing the efficiency of centralised design of transportation networks (in this case the rail networks connecting cities near Tokyo) with *Physarum*-created transport networks approximating the same dataset. The authors found a close correspondence between the two networks in terms of network length and network resilience [42].

Gunji et al. introduced a cellular automaton (CA) model which considered both plasmodial growth and amoeboid movement [55]. The model placed importance on the transformation of hardness/softness at the membrane and the internal transport of 'vacant particles' from outside the membrane resulting in deformation of the original morphology, movement and network adaptation. The model was also able to approximate instances of maze path planning and coarse approximations of the Steiner tree problem and a recent version has been developed to incorporate network growth and adaptation [56]. A hexagonal CA was used by [57] to approximate the growth patterns displayed under differing nutrient concentrations and substrate hardness. The patterns reflected experimental results well but did not exhibit morphological adaptation of the tube network. A recently introduced automata model by Sawa et al. deforms the initial shape of the cell array by exchanging material between cells provided that the difference between two adjacent cells does not exceed a threshold $d$ and the total amount of plasmodium does not exceed another threshold $u$. The model combines features of the Gunji group approach (deformation of an initial mass of plasmodium) with the simple parametric tuning of the Takamatsu approach but as yet does not incorporate response to nutrient sources or growth and shrinkage of the plasmodium [58].

Hickey and Noriega adapted a classical ant colony optimisation algorithm to modify a decision tree in their representation of *Physarum* behaviour in a simple path planning problem [59]. Their algorithm (as with many implementations of ant algorithms) transformed the spatial representation into a graph representation and provided broadly similar results to path optimisation by *Physarum*.

Adamatzky noted that the *Physarum* plasmodium is computationally equivalent, and indeed exceeds the performance of, current prototypes of reaction-diffusion computers [9, 14]. According to Adamatzky the plasmodium *"can be considered as a reaction-diffusion, or an excitable medium encapsulated in an elastic growing membrane."* [60]. The wave propagation of information within plasmodium in response to a complex environment corresponds to a particular type of chemical processor operating in a sub-excitable mode, where the propagation of travelling waves within the plasmodium is influenced by the presence of local nutrient stimuli.

In this paper we give an overview of the multi-agent approach to unconventional computing inspired by slime mould. We try to synthetically reproduce the complex behaviour of slime mould using a simple, particle based, multi-agent model. The aim is to reproduce the complex behaviour of slime mould using equally simple computational components. These simple components interact to produce quasi-material emergent behaviour. The resulting virtual material substrate is, in effect, a 2D material which can be influenced ('programmed') to perform a wide range of spatially implemented computational tasks. We give an overview of the model and the mechanisms which enable computation, before describing a range of computational tasks which it has been used to explore. We conclude with a summary of the approach and an assessment of further avenues of exploration for computation and robotics using this approach.

## 3. Multi-agent Slime Mould Computing by Material Computation

The multi-agent approach to slime mould computing was introduced in [61] consisting of a large population of simple components, mobile agents, (a single agent is shown in Fig. 1) whose collective behaviour was indirectly coupled via a diffusive chemoattractant lattice. Agents sensed the concentration of a hypothetical 'chemical' in the lattice, oriented themselves towards the locally strongest source and deposited the same chemical during forward movement. The collective movement trails spontaneously formed emergent transport networks (Fig. 2) which underwent complex evolution, exhibiting minimisation and cohesion effects under a range of sensory parameter and scale settings. The overall pattern of the population represented the structure of the *Physarum* plasmodium and the individual movement of particles within the pattern represented the flux within the plasmodium. The collective behaved as a virtual material demonstrating characteristic network evolution motifs and minimisation phenomena seen in soap film evolution (for example, the formation of Plateau angles, T1 and T2 relaxation processes and adherence to von Neumann's law [62]). A full exploration of the dynamical patterns were explored in [63] which found that the population could reproduce a wide range of Turing-type reaction-diffusion patterning. The model was extended to include growth and shrinkage in response to environmental stimuli [64, 65]. In a comparison by image analysis and network analysis, the coarsening of the multi-agent networks were found to closely approximate the coarsening observed in the evolution of *P.*

*polycephalum* transport networks [66].

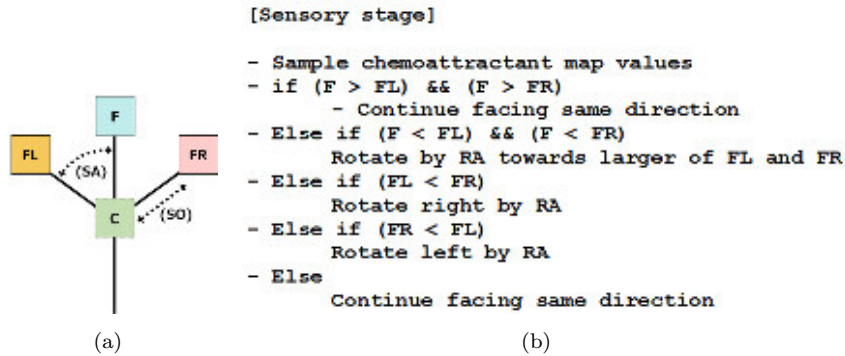

(a)    (b)

Figure 1. Base agent particle morphology and sensory stage algorithm. (a) Illustration of single agent, showing location 'C', offset sensors 'FL','F','FR', Sensor Angle '$SA$' and Sensor Offset '$SO$', (b) simplified sensory algorithm.

## 4. Mechanism of Material Computation

The computational behaviour of the multi-agent approach is generated by the evolution of the virtual material over time and space. Although the type of pattern (and thus the type of material behaviour) can be influenced by parametric adjustment of the $SA$ and $RA$ values (Fig. 3), the evolution is manifested most typically as a shape minimisation over time. At low $SA$ and $RA$ values the patterns are reticulate and adaptive, constantly changing their topology. As $SA$ and $RA$ increase the networks undergo minimisation, reducing the number of edges. Further increases result in labyrinthine patterns and the formation of circular minimal configurations.

The evolution of different patterns, however, does not in itself constitute computation. To perform useful computation we must be able to interact with the material, affecting its behaviour and evolution. This can be achieved by the placement of external attractant and repellent stimuli into the diffusive lattice. These stimuli, when projected into the lattice, form concentration gradients. The gradients constrain the natural minimisation of the material, which would otherwise typically condense the initial inoculation pattern into a minimal configuration. The final (or stable) state of the constrained material indicates that the computation has 'halted'. Because the virtual material is a spatially represented unconventional scheme, the initial problem configuration and final output 'solution' must both be represented as spatial patterns (Fig. 4).

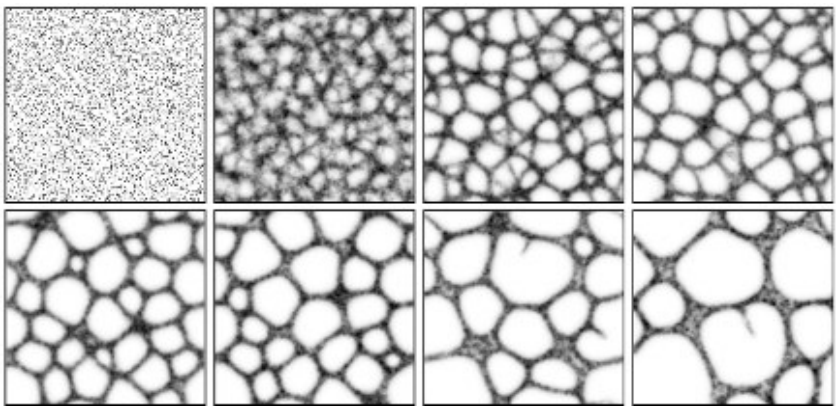

Figure 2. Spontaneous formation and evolution of transport networks in the multi-agent model. Lattice 200×200, %$p$15, $SA$ 22.5°, $RA$ 45°, $SO$ 9, Images taken at: 2, 22, 99, 175, 367, 512, 1740 and 4151 scheduler steps.

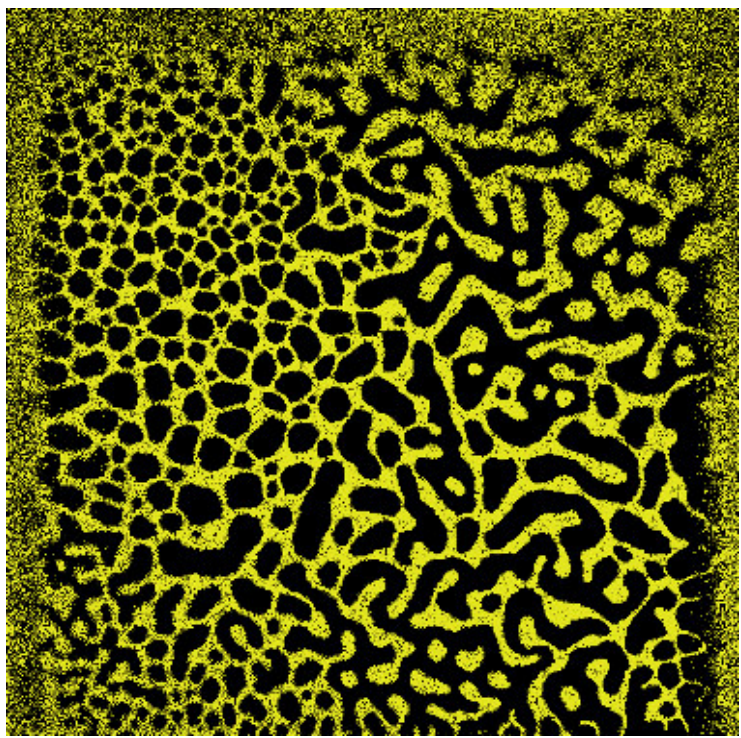

Figure 3.  Parametric mapping of agent sensor parameters $SA$ (across) and $RA$ (down) from $0-180°$ yields complex and dynamical reaction-diffusion patterning. $720 \times 720$ lattice with 181440 particles, $SO = 3$, agents initialised at random positions and orientations, agent distribution at $t = 400$.

The approach follows the scheme suggested by Stepney [67] in which the physical properties of the unconventional computing substrate perform the computation (in this case by material adaptation) and the inputs to the computing substrate can be represented as external fields. In Stepney's scheme there is abundant scope for interfacing the physical substrate with classical computing devices. This interfacing can be used to program the substrate, read the current state of the substrate, extract data from the substrate and to halt the computation when the solution is found. For physical implementations these interfaces include magnetic fields, chemoattractant gradients, light projection (to project input patterns) and video camera systems (to sample current configuration). The interaction between the unconventional substrate and the external system may be very simple, such as the projection of a simple input stimulus pattern, followed by a recording sample of the final state of the computing medium. Alternately, the interaction may be more complex, such as a real-time closed-loop feedback system where the current state of the physical computing medium is sampled, interrogated by a classical computing device and fed back to the substrate. A good example of this more complex integration is the method by Aono and colleagues to control the migration of *Physarum* plasmodium within a stellate chamber, aided by video recording equipment and video projection equipment. Control of input illumination patterns (and thus control of the material substrate) was effected by a Hopfield-type algorithm running on a classical PC [68].

In the specific case of the multi-agent model of slime mould, the computing mechanism can be specified generically in the following scheme (specific implementations and differences will be described in later sections). The multi-agent population resides on a 2D diffusive lattice and the movement of each particle is indirectly coupled to their neighbours by deposition and sensing of chemoattractant within the diffusive lattice. These interactions generate the emergent network formation and minimisation behaviour. This computing substrate is 'programmed' by inoculating the population at specific locations within the substrate, or as a particular pattern (Fig. 5). The evolution of the material is then constrained by placement of external spatial stimuli (chemoattractant gradients, chemorepellent gradients and simulated light irradiation). The computation proceeds with the morphological adaptation of the virtual material (constrained, to some degree, by the stimuli) and the final result is recorded as the stable pattern which the virtual material eventually adopts.

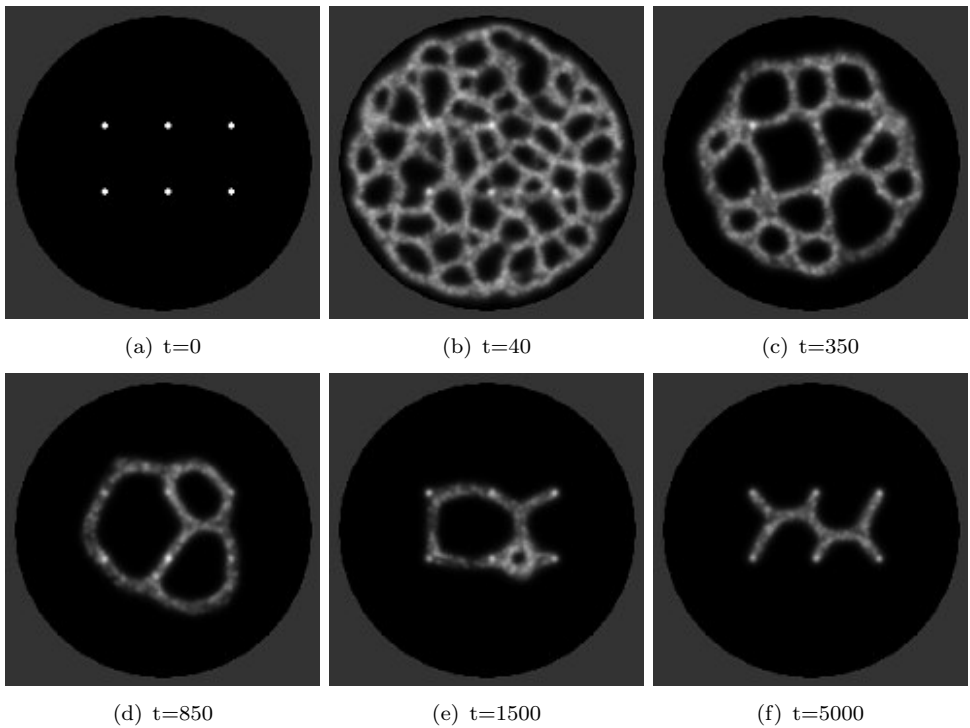

Figure 4. Material computation by constraining pattern formation. a) Pattern of 6 nodes inside circular arena is projected into diffusive lattice as attractant stimuli, b) spontaneous formation of transport network from random inoculation positions (4000 particles, $SA = 60$, $RA = 60$, $SO = 7$), c-e) minimisation of network is constrained by attraction to nodes, f) final stable configuration (output) is a Steiner minimum tree.

## 5. Applications of Multi-agent Material Computation

The adoption of the above generic mechanism for specific computing applications of the multi-agent model of slime mould will be described in this Section. The methods vary in their complexity and in their interactions with classical methods (for example, when acting as a control system or feedback system).

## 6. Graph Problems: Spanning trees and Proximity Graphs

Graph problems are a natural fit for slime mould computing because the growth and adaptation of slime mould naturally forms edges (the protoplasmic tubes of slime mould) between discrete data points (nutrients, such as oat flakes). The global pattern of the slime mould and nutrients can then interpreted as graph edges and nodes. When using living slime mould some allowance needs to be given to the fact that slime mould tubes between nodes are not perfectly straight — they may meander between nutrient nodes significantly. The 'edges' may also not pass through the centre of the nodes exactly. More importantly, there is no such thing as a 'final configuration' of a slime mould graph: the configuration of protoplasmic tubes comprising the slime mould transport network is constantly changing as the organism forages towards new nutrient sources and engulfs and consumes current nutrient sources. During this process, existing tubes may be abandoned (leaving empty remnants of tubes) and growth may sprout from pre-existing tube. Slime mould itself also has to battle with the ever-present threat of competition and predation by other micro-organisms in its computing environment which can affect and disrupt the computation. Nevertheless, graph problems are an ideal candidate to explore slime mould computing.

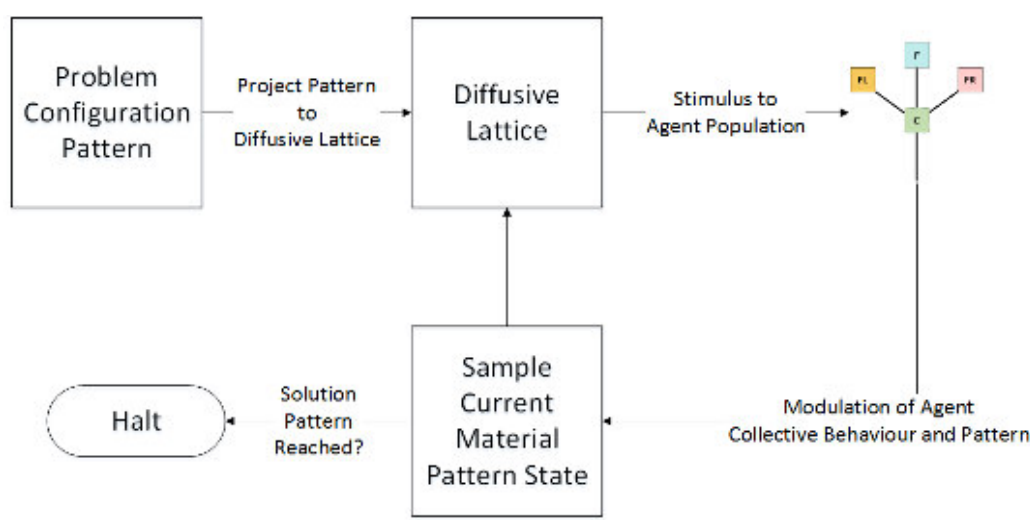

Figure 5. Schematic illustration of unconventional computing approach using multi-agent model of slime mould.

A spanning tree is a connected undirected graph of edges forming a tree structure connecting all of the nodes in a graph without cycles. Proximity graphs are graphs connecting a set of points whose connectivity is determined by particular definitions of neighbourhood and distance [69, 70]. For example, in the construction of the Relative Neighbourhood Graph (RNG), two points $p$ and $q$ are connected only when there is not a third point $r$ that is closer to both $p$ and $q$ than the distance between $p$ and $q$. Each subsequent member of the Toussaint hierarchy of proximity graphs contains the links of the graphs earlier in the hierarchy, adding extra links to satisfy the different neighbourhood definitions. As the hierarchy increases, so does the number of paths and cycles within the graph.

Adamatzky found that a growing plasmodium approximates spanning trees and the Toussaint hierarchy of proximity graphs, noting that current implementations of Belousov-Zhabotinsky (BZ) chemical processors, although able to perform plane division problems such as Voronoi diagrams, are not able to compute spanning trees. This inability of chemical processors is primarily because of the propagation of the wavefronts in all directions from target nodes and also because there is no record left in the medium of the front propagation [14–16].

Slime mould, however, can tackle a wider range of problems than simple chemical processors. In nutrient-poor environments, the plasmodium grows by extending pseudopodia in the specific direction of nearby chemoattractant sources, whereas in nutrient-rich conditions the plasmodium grows radially in all directions (in a manner more akin to classical BZ propagation). Adamatzky suggested that the *Physarum* plasmodium may utilise an efficient method of environmental interaction to guide its behaviour when foraging: When initialised on an oat flake in nutrient-poor conditions, the plasmodium receives local stimuli in the form of a diffusing chemoattractant gradient [16]. The binding of chemoattractant compounds to receptors in the plasmodium modulates the hardness of the plasmodium membrane. The stimulated region softens, and the hydrostatic pressure within the plasmodium vein network causes cellular material to stream towards the source of chemoattractant in the form of a type of pseudopodium (a *lamellopodium*). The plasmodium thus grows towards a nearby node, engulfing it, and stopping (or at least reducing) the diffusion of chemoattractant from the node. The growing pseudopodium then migrates towards the next source of attractant and a spanning tree is ultimately formed. The virtual plasmodium also approximates the construction of spanning trees in nutrient-poor conditions. In the example shown in Fig. 6 A small population was initialised on the lowermost node (Fig. 6a) and grew by extending pseudopodia to link the remaining nodes to form the spanning tree (Fig. 6b-d). Fig. 6e-h shows a representation of the changing chemoattractant gradient field as the model migrates towards and engulfs the nutrient sources.

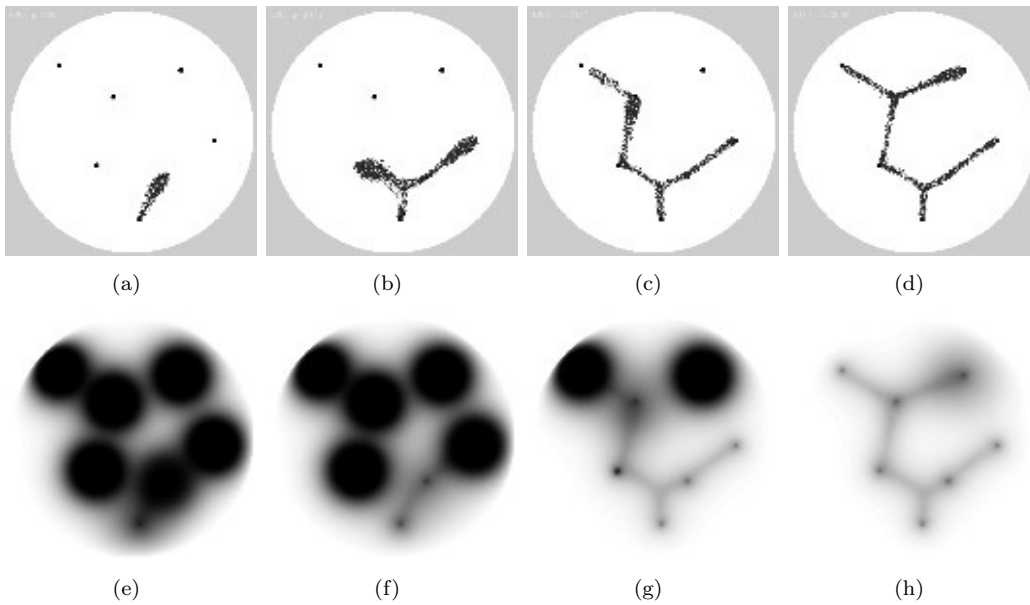

Figure 6. Construction of a spanning tree by the model plasmodium. (a) Small population (particle positions shown) inoculated on lowest node (bottom) growing towards first node and engulfing it, reducing chemoattractant projection, (b-d) Model population grows to nearest sources of chemoattractant completing construction of the spanning tree, (e-h) Visualisation of the changing chemoattractant gradient as the population engulfs and suppresses nutrient diffusion.

When inoculated at a single node *Physarum* was shown to forage to the remaining nodes, approximating a spanning tree. The plasmodium continued its foraging even after the tree was complete, increasing its connectivity to approximate proximity graphs in higher regions of the Toussaint hierarchy (see examples of the hierarchy for simple datasets in Fig. 7), most closely approximating the Relative Neighbourhood Graph (RNG) [16]. When the plasmodium inoculated at all nodes simultaneously, the plasmodium approximated higher connectivity members of the proximity graph family, the Gabriel Graph (GG) and Delaunay Triangulation (DTN), however the connections were thicker between edges corresponding to GG compared to thinner tubes representing DTN edges. The author suggested that the growth front of the plasmodium under different environmental conditions corresponded to the neighbourhood definition of different proximity graphs (for example, the lune neighbourhood of RNG or the circular neighbourhood of GG). Therefore nutrient-poor environments would be expected to induce more sparsely connected networks than nutrient-rich environments (for example, a weak stimulus from only one direction would induce a conical foraging pseudopodium shape, whereas multiple strong stimuli would present a radial growth pattern).

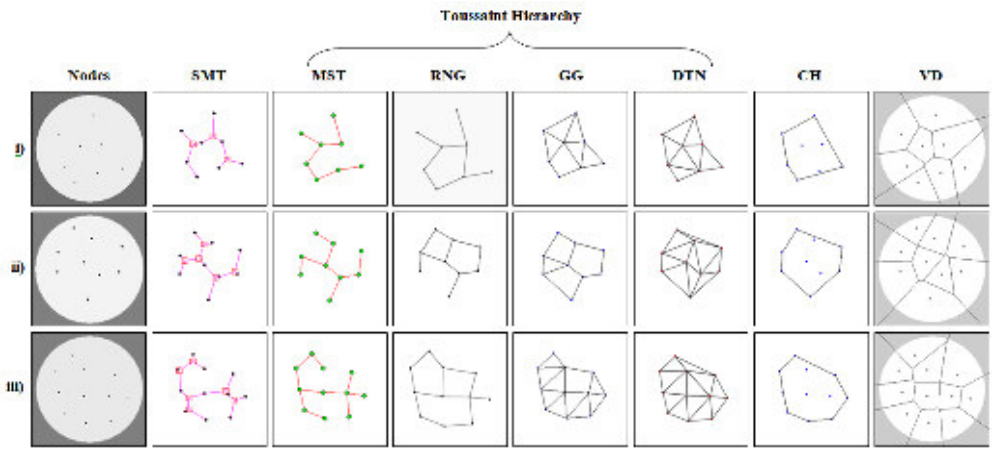

Figure 7. Graphs representing connectivity, proximity, area and tessellation of a set of nodes. Three data sets shown (i–iii), SMT–Steiner Minimum Tree, MST–Minimum Spanning Tree, RNG–Relative Neighbourhood Graph, GG–Gabriel Graph, DTN–Delaunay Triangulation, CH–Convex Hull, VD–Voronoi Diagram.

## 7. Representing Area and Shape: Convex Hull and Concave Hull

Although the proximity graph networks formed by the model connect all of the nutrient sources, they do not group them, provide a representation of the space in which they reside, or provide a representation of their overall shape. This behaviour is shared by slime mould itself. Although slime mould grows outwards from its inoculation site to explore the area of its local environment, any nutrients it discovers are soon connected by its tube network and any direct representation of the overall shape or area is lost. Nevertheless, it is possible to represent area and shape using slime mould using more unorthodox methods. This was achieved by inoculating the plasmodium outside a set of data points containing a substance which acted as a long range attractant and also as a short range repellent (see [17] for more details). The plasmodium propagated around the outside of the data set, approximating the Concave Hull structure. In this section we examine mechanisms to perform these tasks using the multi-agent approach.

### 7.1   The Convex Hull

The Convex Hull of a set of points is the smallest convex polygon enclosing the set, where all points are on the boundary or interior of the polygon (Fig. 8a). Classical algorithms to generate Convex Hulls are often inspired by intuitively inspired methods, such as shrink wrapping an elastic band around the set of points, or rotating calipers around the set of points [71, 72]. Is it possible to approximate the Convex Hull using emergent transport networks by mimicking a physically inspired method? To achieve this we initialise a circular ring of virtual plasmodium *outside* the set of points (Fig. 8b). Because of the innate minimising behaviour of the particle networks the population thus represents a ring of deformable elastic material.

This bounding 'band' then automatically shrinks to encompass the outer region of the set of points. The minimising properties of the paths ensure that the edges of the Hull are straight and convex. There are some practical limitations of this approach. Firstly, the bounds of the set of points must be known in advance, which is not always the case in certain Convex Hull problems. Secondly, points which are inside the final Hull, but close to the 'band' (for example, near the top edge in Fig. 8c) may, via diffusion of their projected attractant, attract the band inwards, forming a concavity. This possibility may be avoided by restricting the nodes to project stimuli only when they have been directly contacted by particles comprising the shrinking band. One benefit of this is that the nodes which are actually part of the final Hull are highlighted (Fig. 8e, the larger nodes).

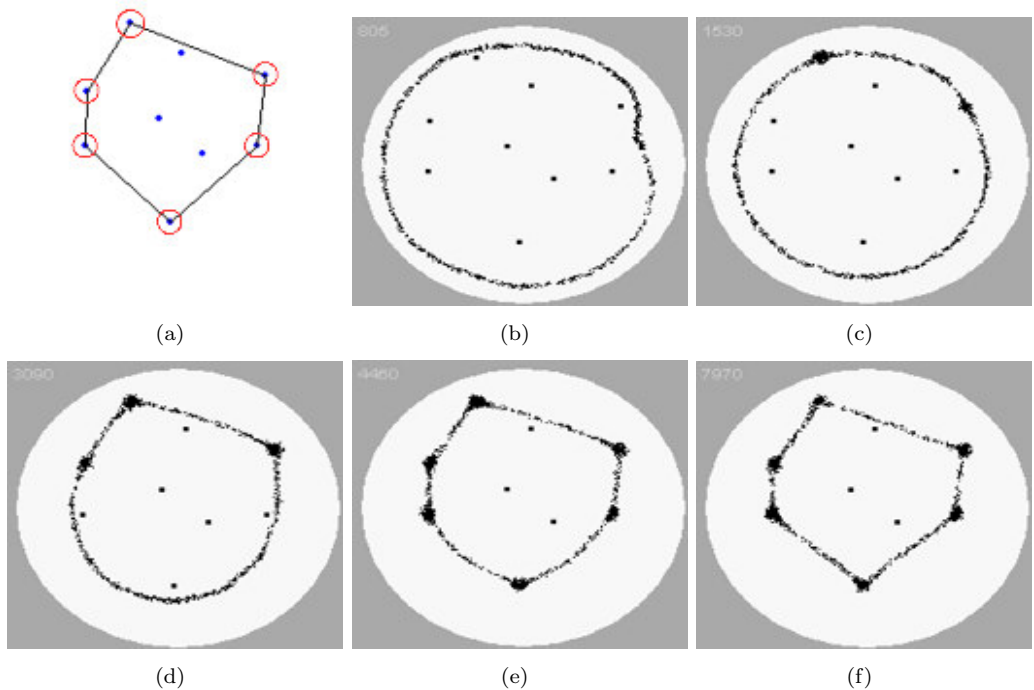

Figure 8. Approximation of Convex Hull by shrinking band of virtual plasmodium. a) original data set with Convex Hull (edges). Nodes which are part of the Convex Hull are circled, b-e) A circular band of virtual plasmodium initialised outside the region of points and shrinks. In this example nodes only emanate nutrients when touched by virtual plasmodia (see text), f) bounding points of final Convex Hull are indicated by larger nodes.

### 7.2 The Concave Hull

The area occupied by, or the 'shape' of, a set of points is not as simple to define as its Convex Hull. It is commonly defined in Geographical Information Systems (GIS) as the Concave Hull, the minimum region (or *footprint* [73]) occupied by a set of points, which cannot, in some cases, be represented correctly by the Convex Hull [74]. For example, a set of points arranged to form the capital letter 'C' would not be correctly represented by the Convex Hull because the gap in the letter would be closed (see Fig. 9a).

Attempts to formalise concave bounding representations of a point set were suggested by Edelsbrunner et al. in the definition of $\alpha$-shapes [75]. The $\alpha$-shape of a set of points, $P$, is an intersection of the complement of all closed discs of radius $1/\alpha$ that includes no points of $P$. An $\alpha$-shape is a Convex Hull when $\alpha \to \infty$. When decreasing $\alpha$, the shapes may shrink, develop holes and become disconnected, collapsing to $P$ when $\alpha \to 0$. A Concave Hull is non-convex polygon representing area occupied by $P$. A Concave Hull is a connected $\alpha$-shape without holes.

The virtual plasmodium approximates the Concave Hull via its innate morphological adaptation as the population size is slowly reduced. A slow reduction in population size prevents hole defects forming in the material which would result in cyclic networks instead of the desired solid shape. The reduction in population size may be implemented by either randomly reducing particles at a low probability rate or by adjusting the growth and shrinkage parameters to bias adaptation towards shrinkage whilst maintaining network connectivity.

In the examples shown below the virtual plasmodium is initialised as a large population (a solid mass) within the confines of a Convex Hull (calculated using the classical algorithmic method) of a set of points (Fig. 9b). By slowly reducing the population size (by biasing the parameters towards shrinkage), the virtual plasmodium adapts its shape as it shrinks. Retention of the mass of particles to the nodes is ensured by chemoattractant projection and as the the population continues to reduce, the shape outlined by the population becomes increasingly concave (Fig. 9c-f).

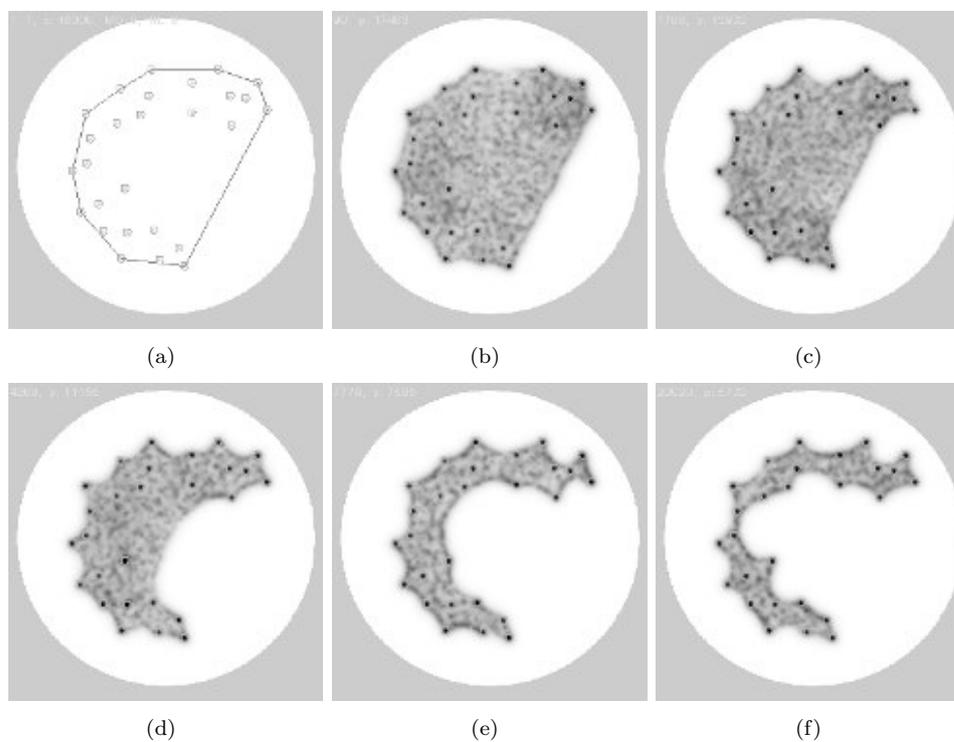

Figure 9. Concave Hull by uniform shrinkage of the virtual plasmodium. (a) Set of points approximating the shape of letter 'C' cannot be intuitively represented by Convex Hull, (b-f) Approximation of concave hull by gradual shrinkage of the virtual plasmodium, $p$=18,000, $SA$ 60°, $RA$ 60°, $SO$ 7.

An alternative approach to shrinkage for constructing a Concave Hull is to grow the structure. It may not be possible to grow the structure when inoculated at a single point source, or at all point sources simultaneously. This is because the individual inoculation sites may be too far apart for the individual particles to sense other regions and fuse. An inoculation pattern must be found which is sparse in representation but which initially spans all the data points. A suitable candidate pattern is the Minimum Spanning Tree (MST) structure of the data points (Fig. 10a and b). This structure guarantees connectivity between all points and also does not possess any cyclic regions. By inoculating the model plasmodium on the MST pattern and biasing the growth/shrinkage parameters towards growth, the model then 'inflates' the MST (Fig. 10c-h) and automatically halts its growth (maintaining a constant population size) as a Concave Hull is approximated. The final image in the sequence compares the grown Concave Hull structure to the Convex Hull of the original point set (Fig. 10h, edges). Note that the grown Concave Hull is a much closer representation of the area of the original point set than the Convex Hull.

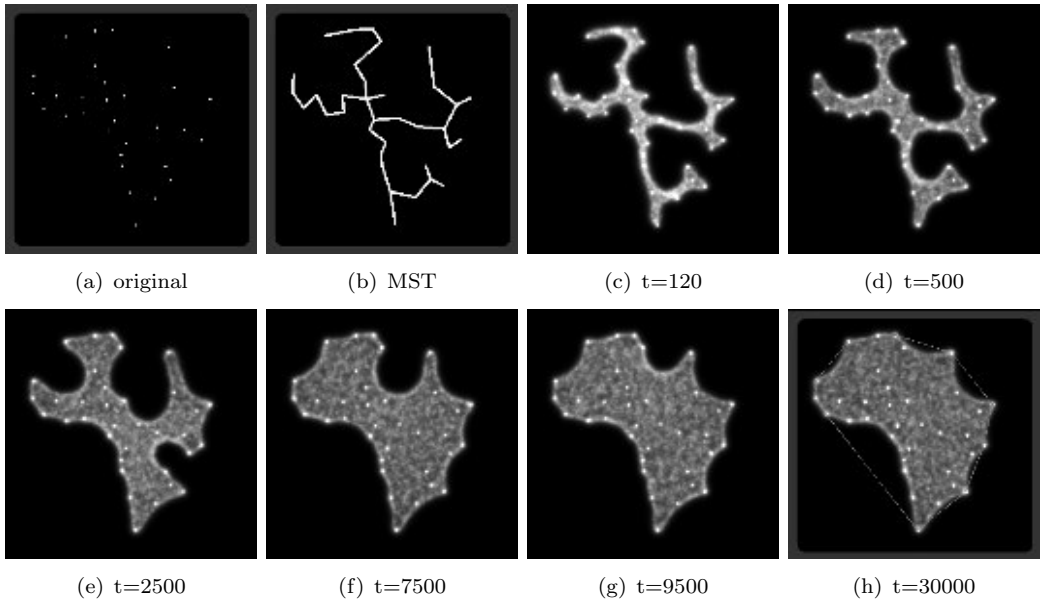

Figure 10. Growth of Concave Hull from Minimum Spanning Tree. a) points representing the locations of major cities in Africa, b) Minimum Spanning Tree of points connects all points without cycles, c-h) after inoculating the virtual plasmodium on the Minimum Spanning Tree the virtual plasmodium grows to approximate the Concave Hull, stabilising its growth automatically (overlaid edges show classical Convex Hull).

It is noteworthy that, in the above example, the population automatically stabilises after 10000 steps. It was shown in [76] that the stabilisation point could be altered by a parameter which effectively tuned the concavity of the final shape of the virtual plasmodium.

## 8. Plane Division: Voronoi Diagrams

If proximity graphs are a natural fit for *Physarum* computing, Voronoi Diagrams, at first glance, would appear to be the opposite. This is because they are structures which divide the plane into discrete areas surrounding the nodes, rather than directly connecting the nodes. The Voronoi diagram of a set of $n$ points in the plane is the subdivision of the plane into $n$ cells so that every location within each cell is closest to the generating point within that cell. Conversely the bisectors forming the diagram are equidistant from the points between them. Voronoi diagrams are useful constructs historically applied in diverse fields as computational geometry, biology, epidemiology, telecommunication networks and materials science. Efficient computation of the Voronoi diagram may be achieved with a number of classical algorithms [77, 78] and are also amongst prototypical applications solved by chemical reaction-diffusion non-classical computing devices [79, 80]. Non-classical approaches are typically based upon the intuitive notion of uniform propagation speed within a medium, emanating from the source nodes. The bisectors of the diagram are formed where the propagating fronts meet, visualised, for example, in chemical processors, by the lack of precipitation where the fronts merge [80]. Voronoi diagrams can be generated in other physical systems by generalising the propagation mechanism and visualisation of the bisectors and have been implemented in a number of different media including reaction-diffusion chemical processors [79, 80], planar silicon [81], crystalline phase change materials [82], and gas discharge systems [83]. In living systems approximation of Voronoi diagrams may be achieved by inoculating a chosen organism or cell type at the source points on a suitable substrate. Outward growth from the inoculation site corresponds to the propagative mechanism and regions where the colonies or cells meet correspond to bisectors of the diagram.

*Physarum* has also been shown to approximate the Voronoi diagram by two different methods, based on its interactions with environments containing repellents and attractants. The method of using *Physarum* to approximate Voronoi diagrams by avoidance of chemorepellents was described in [12, 13]. In this method a fully grown large plasmodium was first formed in a circular arena. Then repellent sources were introduced onto the plasmodium. The circular border of the arena was surrounded by attractants to maintain connectivity of the plasmodium network. The plasmodium then adapted its transport network to avoid the repellents whilst remain connected to the outer attractants, approximating the Voronoi diagram.

Computation of Voronoi diagram may also be achieved by non-repellent methods. This method is proposed in [60] where plasmodia of *Physarum* are inoculated at node sites on a nutrient-rich agar substrate. Attracted by the surrounding stimuli the plasmodia grow outwards in a radial pattern but when two or more plasmodia meet they do not immediately fuse. There is a period where the growth is inhibited (presumably via some component of the plasmodium membrane or slime capsule) and the substrate at these positions is not occupied, approximating the Voronoi diagram. The position of the growth fronts remains stable before complete fusion eventually occurs.

In order to reproduce the formation of Voronoi diagrams with the multi-agent material computation approach, we exploited both attractant and repellent diffusion gradients at varying concentrations [84].

### *8.1 Approximation of Voronoi diagram using repulsion method*

In this method a population of particles was introduced into a circular arena. The interior of the arena was patterned with seven repellent sources and the border of the arena was patterned with attractant sources, reproducing the experimental pattern used in [12]. The uniform distribution of particles was repelled by the field emanating from the repellents and attracted by the sources at the border (Fig. 11), approximating the Voronoi diagram of the repellent sources.

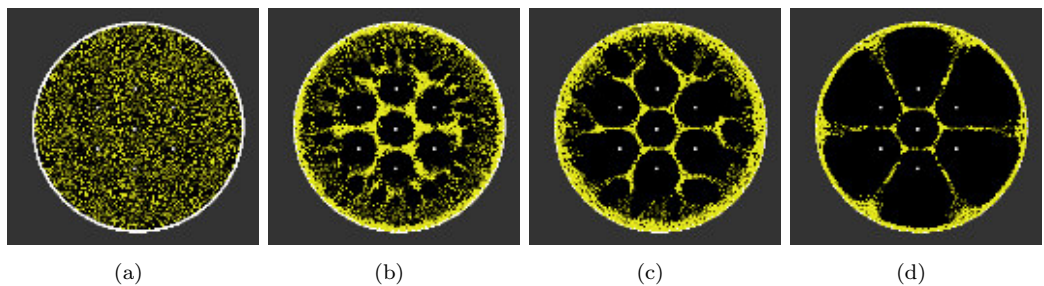

(a)  (b)  (c)  (d)

Figure 11.  Approximation of Voronoi diagram by model in response to repulsive field. (a) Initial distribution of particles (yellow) representing a uniform mass of plasmodium, (b-c) particles respond to repulsive field by moving away from repellents, (d) final network connects outer attractant and bisectors correspond to Voronoi diagram.

## 8.2  *Approximation of Voronoi diagram using merging method*

To approximate the Voronoi by the merging method in [60] a small population of particles were initialised at locations on a simulated nutrient rich background substrate corresponding to Voronoi source nodes (Fig. 12a). The strong stimulation from the high-concentration background generated radial growth (Fig. 12b-c) and the Voronoi Diagram is approximated in the model at regions where the separate growth fronts fuse. Note that the model does not explicitly incorporate the inhibition at the touching growth fronts. In this case the Voronoi bisector position is instead indicated by the model by the increase in network density at the bisector position (Fig. 12d). This approximation is not satisfactory from a computational perspective since it requires subjective visual evaluation and interpretation and the Voronoi bisectors fade after continued adaptation.

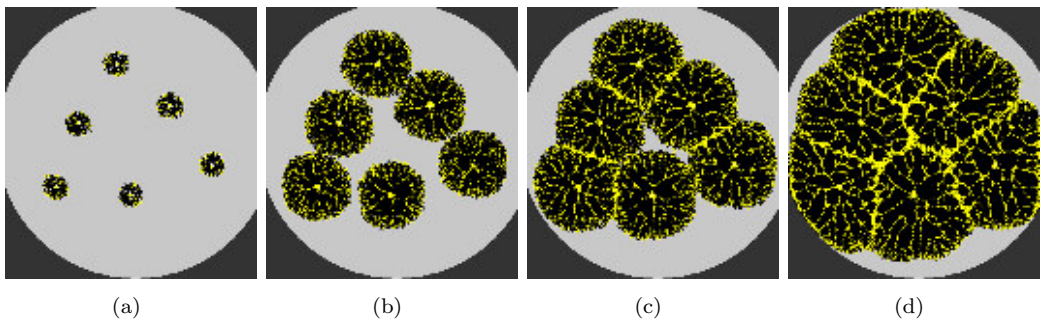

Figure 12.  Approximation of Voronoi diagram by merging method. a-c) expansive growth of separate model plasmodia (yellow) on simulated oat flakes on nutrient-rich background (grey), d) Growth is temporarily inhibited at regions where other model plasmodia are occupied, These dense regions indicates bisectors of Voronoi diagram.

## 8.3  *Hybrid Voronoi Diagrams*

The natural behaviour of the multi-agent population is minimisation of the self-assembled transport networks. By altering the concentration of the repellent field from Voronoi point sources it is possible to generate hybrid Voronoi diagrams with properties of plane division and object wrapping. These structures partition the data points whilst minimising connectivity at the interior of the structure (Fig. 13).

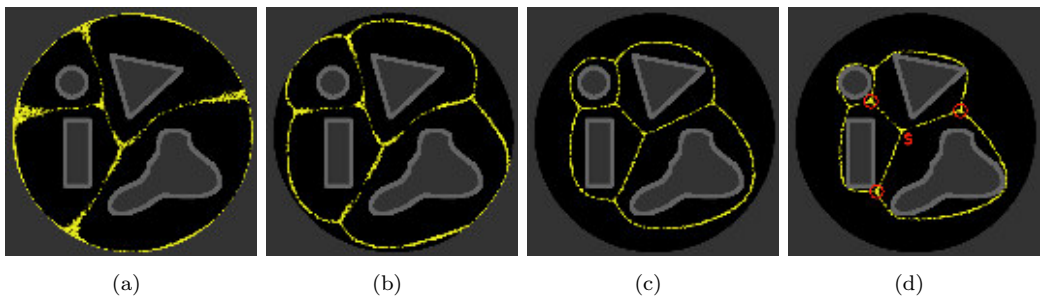

Figure 13.  Reducing repellent concentration allows minimising behaviour to exert its influence, inducing formation of hybrid Voronoi diagram. (a) at high concentration the repellent gradient forces the contractile network (yellow) to conform to the position of curved Voronoi bisectors between planar shapes, (b-d) reduction in repellent concentration allows contractile effects of transport network, minimising the connectivity between cells.

By inoculating small populations on actual stimulus locations it is possible to 'grow' cellular Voronoi diagrams by slowly increasing the repellent concentration.

By utilising different sized stimuli it was possible to approximate weighted Voronoi diagrams. Smaller stimuli generated smaller repellent gradient fields whereas larger stimuli generated fields which exerted their repellent effect over a larger distance (Fig. 14).

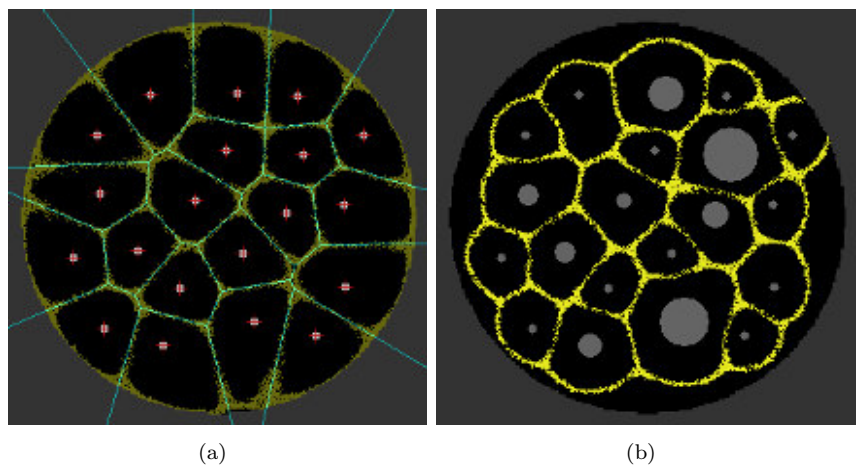

Figure 14. Comparison of conventional and weighted Voronoi diagram. a) multi-agent model (yellow) approximates the Voronoi diagram (blue) at high -ve stimuli concentration (Voronoi diagram by classical method is overlaid), (d) Weighted diagram is approximated by varying size of source data points.

## 9. Path Planning

The examples provided so far concern graph structures comprised of data points and edges, and the networks and shapes they represent. Another problem that can be tackled by the material computation approach is the problem of how to find a path between two points in an open arena. This is a problem typical in robotics applications. An approach was devised in [85] whereby a large population of the virtual plasmodium was inoculated within a lattice whose boundaries took the shape of a robotic arena. Chemoattractant was projected into the lattice at the desired start and end locations and the population size was reduced. The model plasmodium adapted its shape automatically as the population size reduced, conforming to the boundaries of the arena and forming a path between the start and end locations.

A number of variations on the approach are possible. To ensure collision-free paths (e.g. paths which are not too close to walls) it is possible to utilise diffusion of repellents from the arena walls. This resulted in paths which took a more central trajectory through the arena. To prevent multiple paths forming around any obstacles in the arena a method was devised that projected chemorepellent into the lattice at sites where obstacles were located only when they were partially uncovered by the shrinking blob (see Fig. 15 for an example). The repellent field pushed the blob away from the obstacles, ensuring that only a single path from the start to exit was formed.

## 10. Combinatorial Optimisation

The Travelling Salesman Problem (TSP) is a combinatorial optimisation problem well studied in computer science, operations research and mathematics. In the most famous variant of the problem a hypothetical salesman has to visit a number of cities, visiting each city only once, before ending the journey at the original starting city. The shortest path, or tour, of cities, amongst all possible tours is the solution to the problem. The problem is of particular interest since the number of candidate solutions increases greatly as $n$, the number of cities, increases. The number of possible tours can be stated as $(n-1)!/2$ which, for large numbers of $n$, renders assessment of every possible candidate tour computationally intractable. Besides being of theoretical interest, efficient solutions to the TSP have practical applications such as in vehicle routing, tool path length minimisation, and efficient warehouse storage and retrieval.

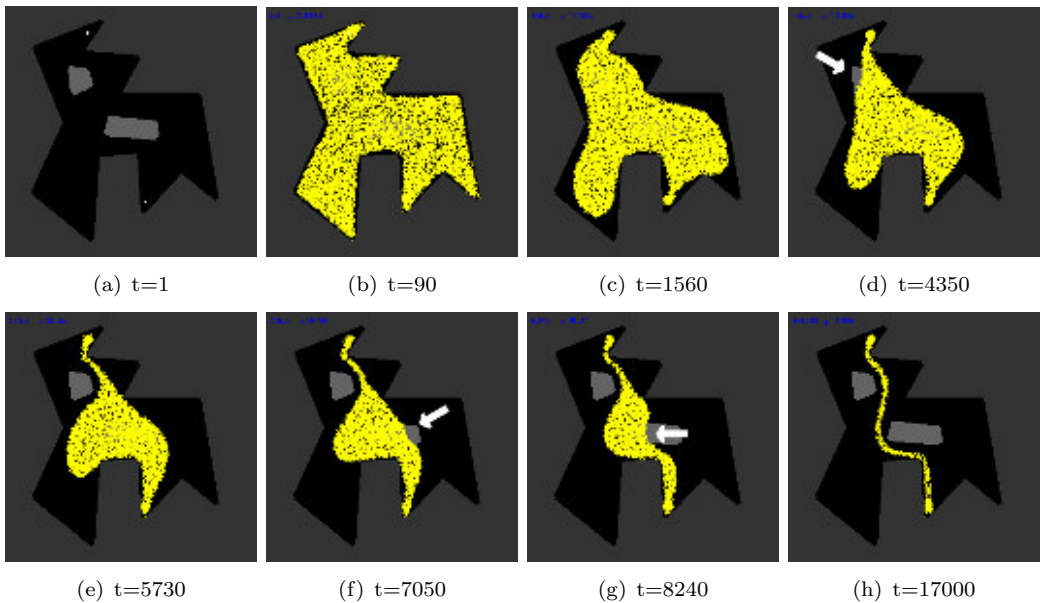

(a) t=1  (b) t=90  (c) t=1560  (d) t=4350

(e) t=5730  (f) t=7050  (g) t=8240  (h) t=17000

Figure 15. Path planning by shrinkage and morphological adaptation. a) arena with habitable areas (black), inhabitable areas (dark grey), obstacles light grey and path source locations (white). b) blob initialised on entire arena, including obstacles, c) gradual shrinkage of blob, d) exposure of obstacle fragment generates repellent field at exposed areas (arrow), e) blob moves away from repellent field of obstacle, f) lower obstacle is exposed causing repellent field at these locations (arrow), g) further exposure causes migration of blob away from these regions (arrow), h) final single path connects source points whilst avoiding obstacles.

The intractable nature of the TSP has led to the development of a number of heuristic approaches which can produce very short — but not guaranteed minimal — tours. A number of heuristic approaches are inspired by mechanisms seen in natural and biological systems. These methods attempt to efficiently traverse the candidate search space whilst avoiding only locally minimal solutions and include neural network approaches (most famously in [86]), evolutionary algorithms [87], simulated annealing methods [88], the elastic network approaches prompted in [89], ant colony optimisation [90], living [68] and virtual [91] slime mould based approaches, and bumblebee foraging [92]. The multi-agent model was applied to the TSP using two different methods, differing in their complexity.

### 10.1 Dynamic Reconfiguration of Transport Network Method

In the first approach [91] a method was devised to try and dynamically control in real-time the evolution of the emergent multi-agent transport networks. This required a mechanism to analyse the current connectivity state of the transport network generated by the agent population. This analysis generated a map of connectivity at each node of the network, in terms of node degree (the number of other nodes to which each node is connected). Given this node connectivity information a series was rules was applied to each node in turn. If a node had $d > 2$ the concentration of attractant projected at that node was reduced. If a node had $d <= 2$ the concentration of attractant was increased. A reduction in attractant concentration reduced the attraction of the network to that node, causing it to detach from the network. Conversely, an increase in attractant increased the attraction of the network to that node, causing it to become attached to the network and increase the tension of the material in this location.

The resulting network evolution was extremely complex (an example is shown in Fig. 16). Characteristic high-level motifs of network changes were observed as the network moved from one configuration to another. In simple networks the evolution was guided to minimal TSP configurations where $d = 2$. However, the method was 'blind' in that there was no guarantee that a solution would be found and, if one were found, there was no guarantee that the network would represent a valid TSP tour.

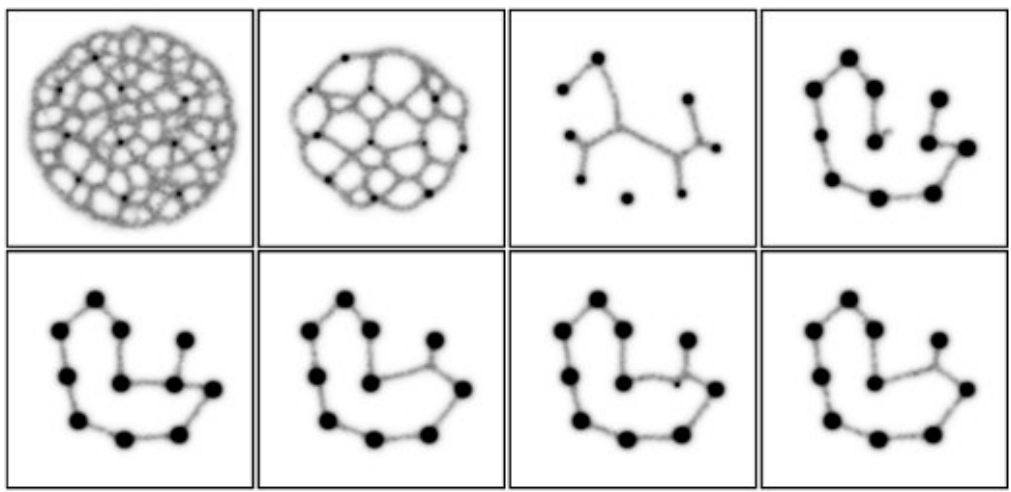

Figure 16. Dynamical reconfiguration of multi-agent transport network using feedback control. Initial random network (grey lines) formed (top left) and dynamical evolution of network influenced by changing node concentrations (dark grey spots).

## 10.2 Shrinking Blob Method

A simpler approach to combinatorial optimisation was later used in which morphological adaptation of a solid 'blob' of virtual slime mould was used to generate solutions to the TSP [93]. In this method the multi-agent population was initialised in the shape of a Convex Hull covering the set of nodes comprising TSP city locations (Fig. 17a). Each city location projected a source of attractant into the diffusive lattice but this projection was damped if the city location was completely covered by the mass of the blob. The population size was then systematically reduced in size over time. As the blob shrunk it partially uncovered cities which were underneath the blob. As each city was uncovered its projection strength was increased, causing the blob to adhere to this new location. The shrinking blob automatically adapted its shape to conform to the new stimuli (Fig. 17b-e) and the shrinkage process was halted when all cities were uncovered. This spatially represented approach also displays a spatially represented *solution* — a tour which can be interpreted by tracing the periphery of the final blob shape (Fig. 17f). The method constructs TSP tours by incrementally inserting new cities into the initial city list found in the Convex Hull structure. The city list then grows as the blob becomes more concave. The method is also notable in that only a single tour is generated in each run with no further attempts at reducing the tour distance. Nevertheless, tours discovered using this simple method were found to be between only 4% and 9% longer than optimal tours computed by brute force methods.

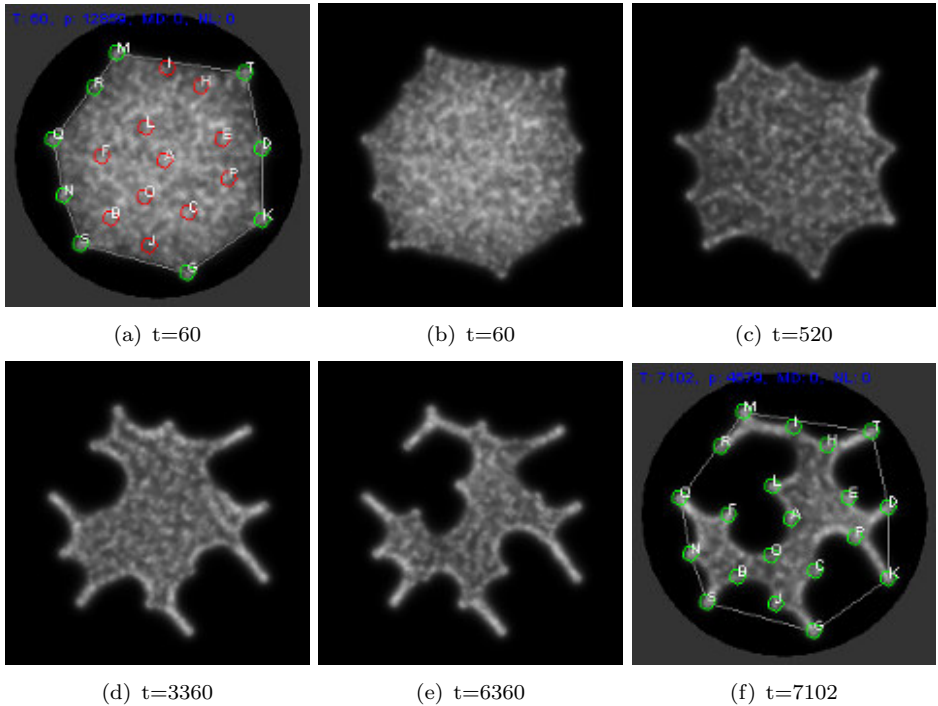

Figure 17. Visualisation of the shrinking blob method. (a) sheet of virtual material initialised within the confines of the convex hull (grey polygon) of a set of points. Node positions are indicated by circles. Outer partially uncovered nodes are light grey, inner nodes covered by the sheet are in dark grey, (b-e) sheet morphology during shrinkage at time 60, 520, 3360, 6360 respectively, (f) shrinkage is stopped automatically when all nodes are partially uncovered at time 7102.

## 11. Data Smoothing and Filtering

Smoothing and filtering plays an important role in signal processing. In [94] we assessed if the innate morphological adaptation behaviour of the *Physarum* model could be used for data smoothing applications. We represented a 1D signal as a sequence of Y-axis data values along a time-line represented by X-axis values (Fig. 18a). The virtual material was patterned in this initial shape and was initially held in place by projecting attractants in the shape of the original pattern (Fig. 18b). When the attractant stimulus was removed the material relaxed and adapted into a profile which smoothed the data, matching the moving average of the original data. The moving average is computed conventionally by a kernel computing the mean of the current data point and its left and right data points of window size $w/2$. The moving average filter has non-periodic boundary conditions and subsequently the moving average line narrows as the kernel window size increases (to prevent the window exceeding the bounds of the data). The amount of smoothing by the material deformation (i.e. corresponding to the kernel window size of the moving average) was dependent on the length of time that the material adapted for (Fig. 18c-f). Note that the width of the material also shrinks over time, mirroring the narrowing of the moving average line as the kernel window increases.

Instead of completely removing the original data stimuli it is possible to use the original stimuli to constrain the adaptation of the model. This was achieved by projecting a weaker representation of the original data stimuli into the lattice. The effect of maintaining a weakened stimulus strength was subtly different to the moving average and appeared to partially filter the data. Specifically, the material tended to adhere in areas of the data that were relatively unchanging, whilst detaching from regions that underwent large changes in direction (Fig. 19). This behaviour corresponds to that of an iterative low-pass filtering process where high frequency signal components are removed whilst low frequency components remain. Note that by maintaining a weak stimulus, the width of the material spanning the dataset was not significantly reduced, in contrast to the material approximation of the moving average.

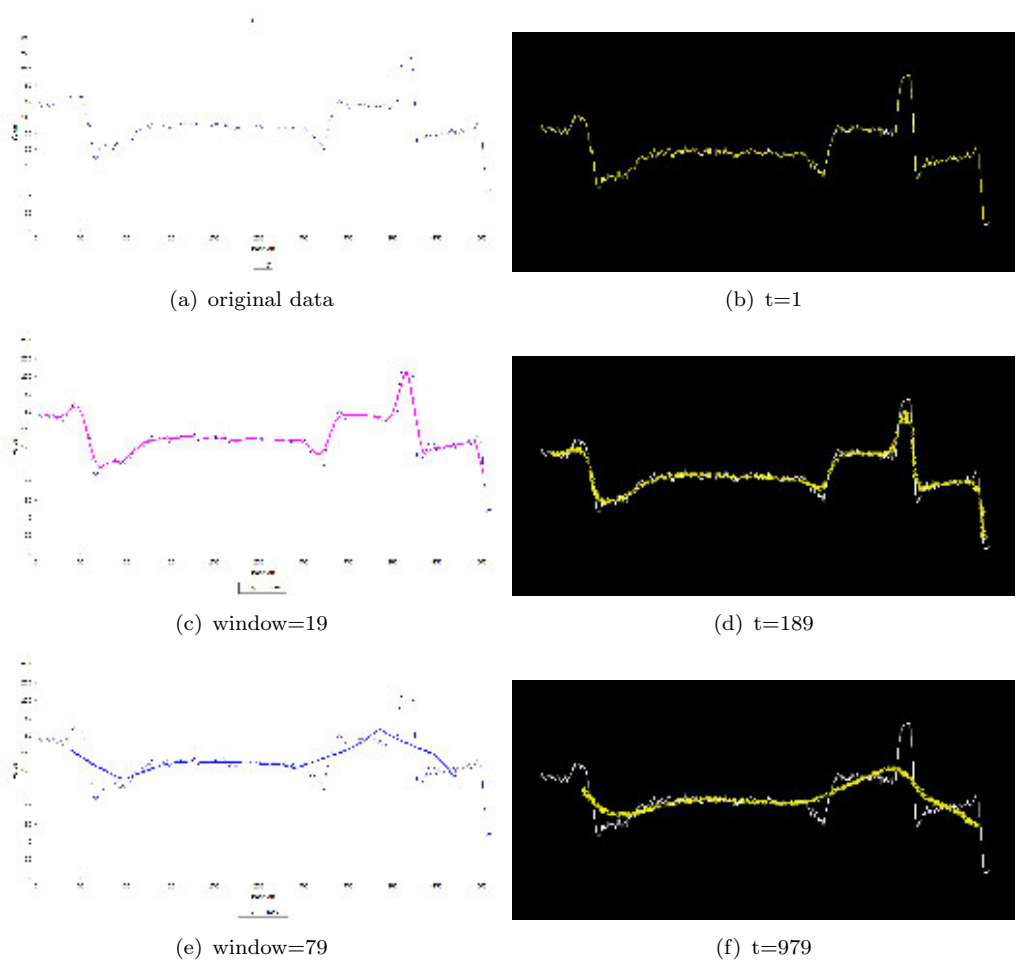

Figure 18. Relaxation of virtual material approximates the moving average. (a,c,e) original data (thin line) and overlaid moving average filtered data (thick line) with 1D kernel of size 19, and 79 respectively, (b,d,f) initialisation of virtual material on original data followed by snapshots at increasing time intervals.

## 12. Approximation of Spline Curves

Splines are mathematical functions constructed piecewise from polynomial functions. Spline functions connect separate data points with a smooth continuous curvature where the individual functions join (at regions called *knots* [95],[96]. Spline functions are useful for curve fitting problems (*approximating splines*, where the spline smooths the path between data points) [97]. They may also be used in interpolation problems (*interpolating splines*, where the spline curve passes through all of the data points) [98]. Due to the natural curvature enabled by spline curves, and rapid development in computer aided design systems, they have proven popular in design and architecture [99]. The term spline apparently refers to the use of flexible strips formerly used in the shipbuilding and motor-vehicle industries to allow the shaping of wood and metal shapes into smooth forms by deforming them at selected points using weighted metal objects known as ducks. Thus, there is an inherent mechanical nature to the operation and interpretation of spline curves. The mechanical properties have been used as an inspiration for deformable models and templates, initiated by [100], primarily for image segmentation, and subsequently extended to 3D application [101], [102]. We assessed the potential for approximating spline curves using the collective and emergent material shrinkage in the multi-agent model [94].

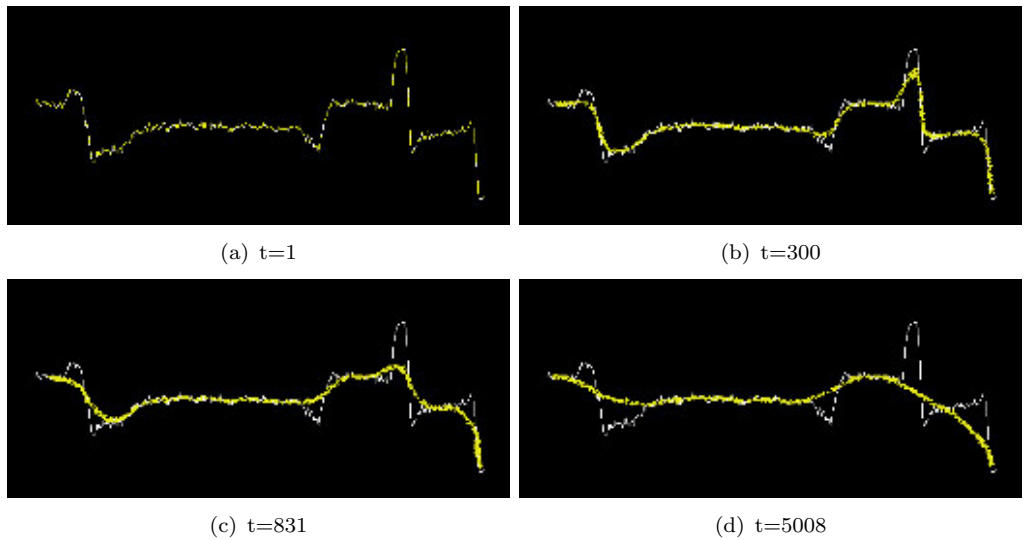

(a) t=1            (b) t=300

(c) t=831            (d) t=5008

Figure 19. Weak background stimulus constrains adaptation of the virtual material approximating a low pass filter. a) Initialisation of virtual material on original data weakly projected into the lattice at concentration of 0.255 units, b-d) snapshots at increasing time intervals showing removal of sharp peaks and troughs in the data.

Fig. 20 shows a set of 20 points which was used to generate a B-spline curve of degree 2 and 5 using numerical methods. The virtual material was initialised in the path of the original polyline and held in place by a weak attractant stimulus in the pattern of the polyline connecting the data points. The material relaxed over time when the initial attractant stimulus was removed (attractant was retained at the two end points to clamp the relaxing material). The morphological adaptation of the multi-agent population approximated the B-spline curve. Increasing the relaxation time period corresponded to increasing degree of the original spline curve (Fig. 20 c,d).

Closed shapes may be represented by clamped spline curves by repeating the same start data points. For unclamped open shapes, overlapping the first and last three points generates a smooth open curve (Fig. 21,a). For clamped open shapes the first point is overlapped by the end point (Fig. 21,b). For the material approximation of open spline curves the material is simply patterned with the closed polyline (Fig. 21,c) and if a clamping point is required this is represented by attractant projection at the desired clamping site (Fig. 21,d).

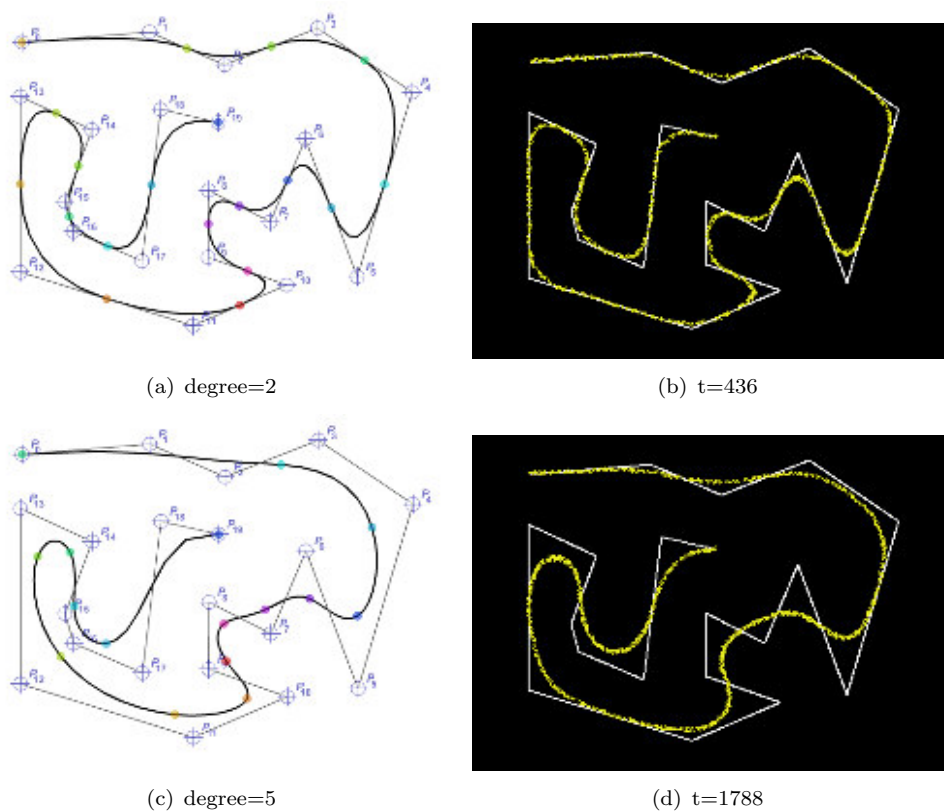

(a) degree=2      (b) t=436

(c) degree=5      (d) t=1788

Figure 20. Clamped B-spline of complex shape and approximation by virtual material. Left: B-spline at differing degrees, composed of 20 points, clamped at start and end points. Points shown as labelled hollow circles connected by faint lines, knots shown as solid circles on thicker spline curve. Right: Approximation of spline curve by relaxation of virtual material, clamped at start and end points.

## 13. Spatial Approximation of Statistical Properties

The sclerotium stage is a part of the life cycle of slime mould, whose entry is provoked by adverse environmental conditions, particularly by a gradual reduction in humidity. In prolonged dry conditions the mass of plasmodium aggregates together, abandoning its protoplasmic tube network to form a compact, typically circular or elliptical, toughened mass [103]. Sclerotinisation protects the organism from environmental damage and the slime mould can survive for many months — or even years — in this dormant stage, re-entering the plasmodium stage when moist conditions return. Biologically, the sclerotium stage may be interpreted as a primitive survival strategy and it has been interpreted computationally as a biological equivalent of freezing or halting a computation [104] in spatially represented biological computing schemes.

In [105] we investigated whether sclerotinisation could inspire a method of spatial computation whereby useful statistical information about a spatially represented dataset could be summarised by morphological adaptation. We first chose the task of centroid computation of 2D shapes. The geometric centroid is a weighted mean of all the X and Y co-ordinates of a shape. For a two-dimensional shape with uniform thickness the centroid can be considered as the centre of mass of the shape and, for certain complex shapes, the centre of mass may even lie outside the shape itself.

To assess different morphological adaptation methods and to see how well the adaptation approximates the centroid we initiated a large mass of virtual material in the pattern of a number of shapes. The shapes selected have different properties, such as solid, containing holes, concave, and convex. The material was held in the initial pattern by projecting attractants into the lattice corresponding to the original pattern for 50 scheduler steps. The centroid of each of the original patterns was computed conventionally by the mean value of all points within the pattern (for example, Fig. 22a, circled). Since the particle population was initially configured as the original pattern the centroid of the population obviously initially matched the centroid of the original. The attractant stimuli was then completely removed from the lattice and the material underwent morphological adaptation via its emergent relaxation behaviour. The population was reduced in size by randomly removing particles from the blob (at probability $p =0.0005$). As particles were removed the blob automatically shrunk in size, the shrinkage of the blob allowing a visual result of the centroid position (Fig. 22a-f). The centroid of the virtual material was computed conventionally by averaging the co-ordinates of all particles within the blob and compared to the centroid of the original pattern. The experiments were halted when the population size became <50. The Euclidean distance between the original centroid and blob centroid (the mean absolute error, MAE) over ten runs is shown in the graph in Fig. 22g.

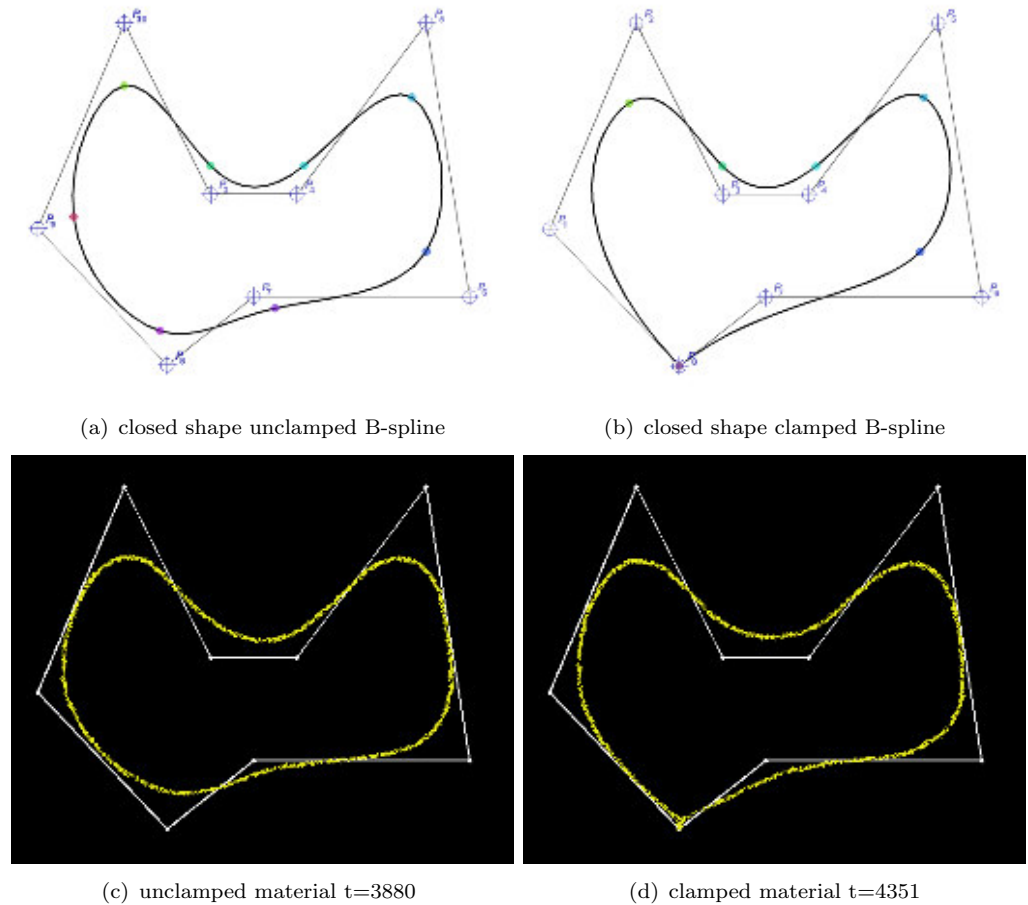

(a) closed shape unclamped B-spline  (b) closed shape clamped B-spline

(c) unclamped material t=3880  (d) clamped material t=4351

Figure 21. Approximation of B-spline curves in closed cyclic shapes. a) closed shape with no clamping points, b) closed shape clamped at point 8, c) material approximation of unclamped shape, d) material approximation of clamped shape by projecting attractant at data point 8.

The results for the lizard shape indicate that as the material adapts to the removal of stimuli and the shrinkage process, it is able to approximate and maintain the same centroid position as the original shape to within - on average - two pixels accuracy. At the start of the process the error accumulates but stabilises after 6000 steps (MAE 2.09, $\sigma$ 1.36). Results for a variety of 2D shapes are shown in Fig.23. Each sub-figure shows the centroid position of the original shape (marked

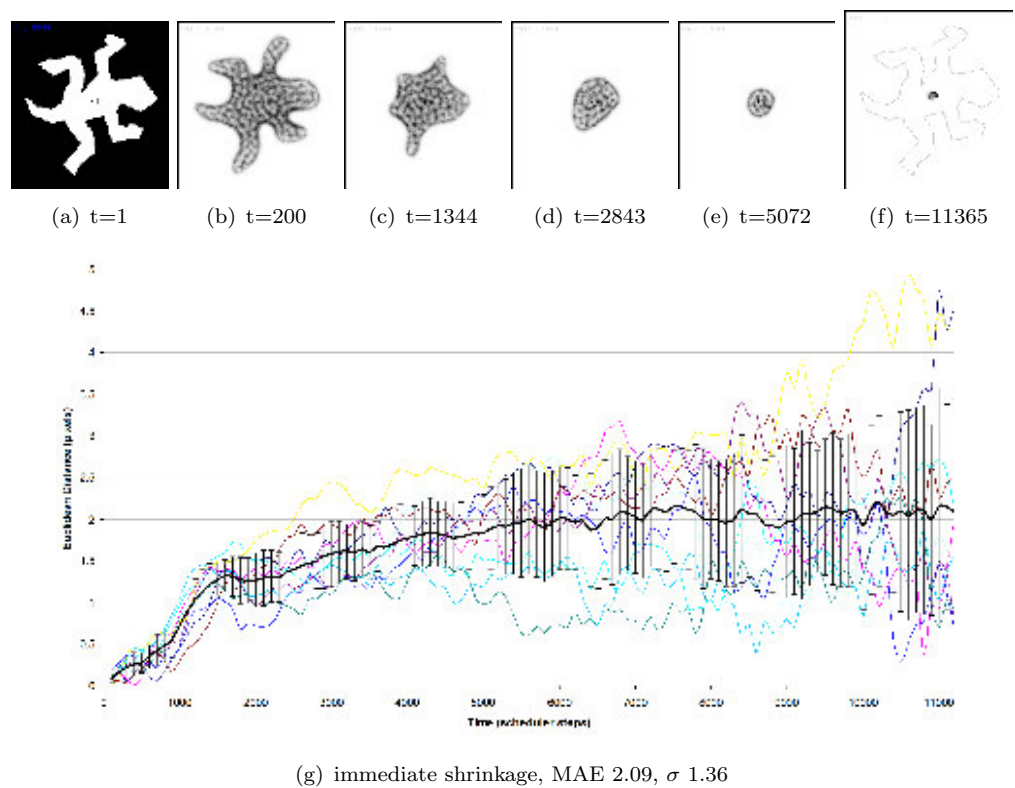

(g) immediate shrinkage, MAE 2.09, $\sigma$ 1.36

Figure 22. Approximating the centroid by morphological adaptation and shrinkage. a) initial lizard shape with centroid indicated (circle), b-f) After initialisation in the original pattern the material undergoes adaptation, shrinking and relaxing to approximate a circular shape, g) chart plotting mean absolute error of blob centroid from original image centroid during adaptation with simultaneous shrinkage. 10 runs are shown overlaid (faint lines) with mean (thick line) and standard deviation error bars.

by a red cross, online) and the distribution of blob centroids (the final position of the blob after adaptation and shrinkage) over ten runs, marked as a distribution blue dots (online). The results show better performance at tracking the centroid of convex shapes, including those with holes (Fig.23a-c). As shapes become increasingly concave, the error begins to increase. The worst performance is on shapes with strongly concave features where the centroid lies outside the boundary of the original shape (Fig.23e-f).

In addition to summarising the properties of 2D shapes, can we utilise the shrinkage method to summarise arithmetic properties of a spatially represented numerical dataset? To assess this possibility we randomly generated 20 numbers from the range of 0 to 100 and used these values as Y axis positions. The generated data values were sampled from a uniform distribution but had a wide range of variance across all experiments ($\sigma$ of between 542 and 1293 for the sorted data experiments and $\sigma$ between 462 and 1159 for the unsorted data experiments). X axis positions were generated using regular spacing of 20 pixels between the data points and we then connected these data points to give a shaped path on which to initialise the virtual material. The method was assessed over 50 randomly generated datasets for both unsorted lists of data (Fig. 24) and for data points pre-sorted by value (not shown). During each run the virtual material was initially held in place by attractant projection for 20 steps of the model (Fig. 24a) and the attractant was then removed, causing the adaptive population to smooth and shrink the original shape. For unsorted data values the initial behaviour was to shrink away from sharp peaks and troughs connecting the data points (Fig. 24b) until an approximately smooth line was formed (Fig. 24c). This band of material then shrunk horizontally from each end (Fig. 24d,e). Each experiment was halted when the population size of the blob was $< 50$ and the final Y-axis position of the centre of the remaining population was compared to the arithmetic mean of the original data (hollow circle in Fig. 24f). In the case of the pre-sorted data values the smoothing of the line was much more short lived and the line began shrinking from both ends almost immediately. For the unsorted data points the mean error of the final blob position when compared to the numerically calculated arithmetic mean was 5.90 pixels ($\sigma$ 3.7) and for the pre-sorted population the mean error was 2.23 pixels ($\sigma$ 1.72). We did not find any strong correlation between the standard deviation of the randomly generated data points and performance (error) of the final position of the blob (Pearson correlation coefficient of $\rho = $ -0.07 for sorted data and $\rho = 0.09$ for unsorted data).

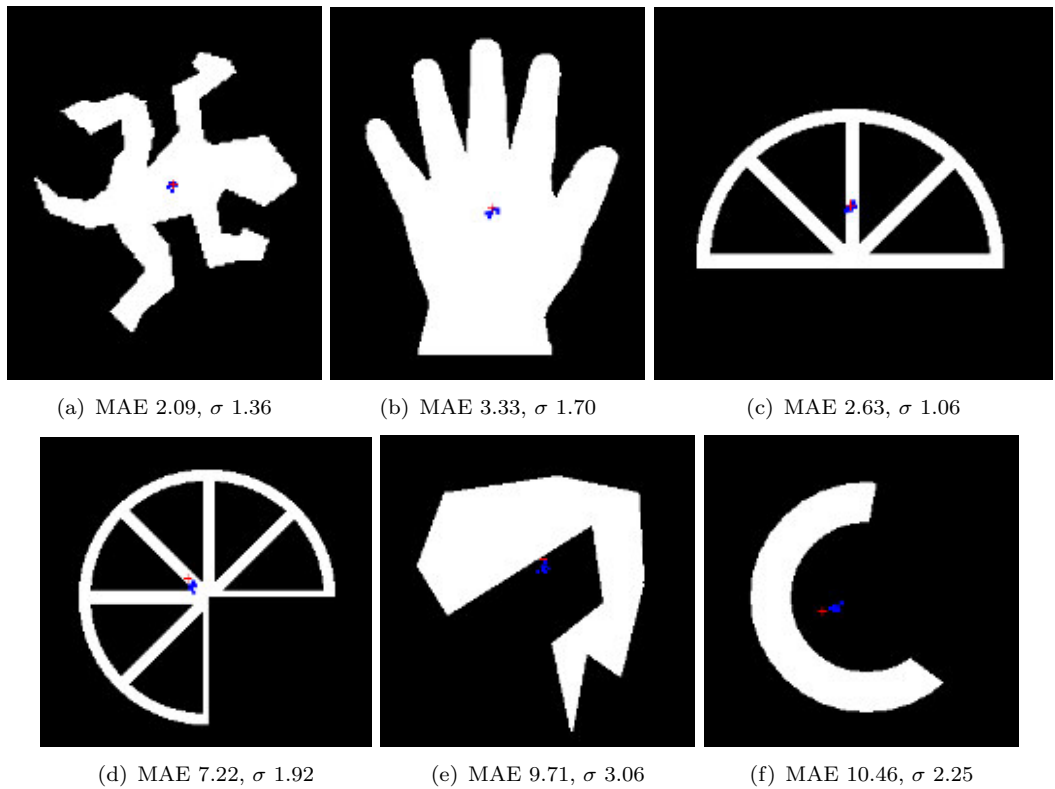

(a) MAE 2.09, $\sigma$ 1.36  (b) MAE 3.33, $\sigma$ 1.70  (c) MAE 2.63, $\sigma$ 1.06

(d) MAE 7.22, $\sigma$ 1.92  (e) MAE 9.71, $\sigma$ 3.06  (f) MAE 10.46, $\sigma$ 2.25

Figure 23. Illustration of difference between blob approximation of centroid and image centroid. a-f) original image with centroid position (red cross shape, online) and distribution of blob positions over 10 runs (blue dots, online) with mean absolute error (MAE) and standard deviation ($\sigma$) indicated in labels.

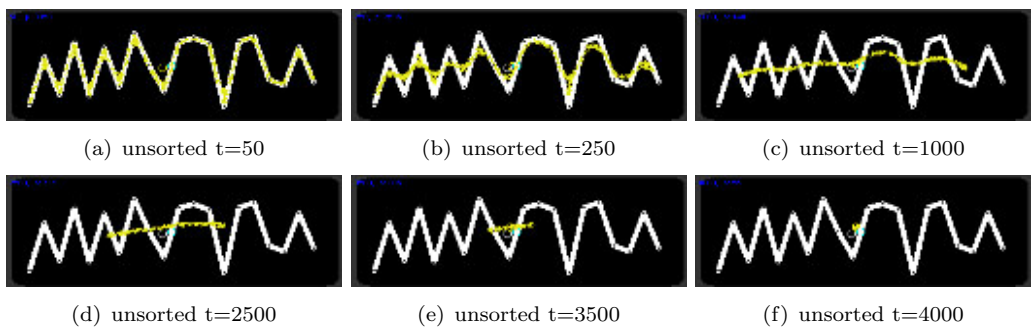

(a) unsorted t=50  (b) unsorted t=250  (c) unsorted t=1000

(d) unsorted t=2500  (e) unsorted t=3500  (f) unsorted t=4000

Figure 24. Approximation of arithmetic mean of 1D data by morphological adaptation. Individual data points indicated on inverted Y-axis by dark dots on connected line, adaptive population shown as coarse shrinking blob.

## 14. Conclusions

We have described a multi-agent model of slime mould *Physarum polycephalum* which was used as a spatially represented unconventional computing substrate. As with slime mould itself, the individual componentes of the model have very simple behaviours and computation is an emergent property of their collective interactions. The model behaves as a virtual material which, like *Physarum*, can be 'programmed' by the placement of stimuli — attractants and repellents. Computation occurs in the model by constraining the natural pattern formation of the virtual material in response to the diffusive gradients of the stimuli. We have presented examples of unconventional computing with the model where problems must be represented as spatial patterns. The constrained patterning then performs the computation itself (occasionally aided by interaction with classical PC devices), and the final solution of the problem is also represented by a stable spatial pattern.

Example applications have been demonstrated for a range of tasks. All of the tasks exploit the material relaxation phenomena of the model. Some tasks are a natural fit to slime mould computing, such as the formation and evolution of spanning trees and proximity graphs. Other tasks exploit the response to repellents for plane division, whilst hybrid structures can be approximated by combining the attractant contraction response with the repellent avoidance response. The overall shape of a dataset was approximated by mimicking wrapping approaches to the Convex Hull, whilst the Concave Hull was approximated by growing the virtual material from a spanning tree. The path planning task was performed by a shrinking sheet of virtual material, adhering to attractant sources and performing obstacle avoidance via repellent fields from obstacles and walls. The morphological adaptation of the multi-agent model was used to approximate low-pass filters and spline curves. It was also found that a shrinking mass of the model plasmodium, when mimicking the sclerotinisation response of the slime mould, could approximate the Centroid of a complex shape and approximate the mean of a sequence of spatially represented numerical data.

More complex interactions with the model were used for combinatorial optimisation tasks. A realtime network analysis and feedback method was devised to indirectly guide the connectivity of the multi-agent transport networks by dynamically adjusting attractant weights. In a simpler approach, the Travelling Salesman Problem was performed by a shrinking mass of model plasmodium and the shrinkage was automatically halted when all cities were added to the tour.

Some particular challenges to using spatially implemented unconventional computation still remain. What are the problems that can — and cannot — be performed by these methods? Are some problems more suited than others? How easy is it to implement these problems as spatial patterns? In classical programming there may be multiple ways to encode the solution to a problem, does the same apply to spatially represented problems?

How do we know when the computation of the problems has finished? In some of the examples in this paper, we interacted with the material using a classical PC program to record and analyse the progress of the computation. Alternatively, this may be performed by detection of when a stable pattern state is reached. This can be performed by relatively simple image analysis. Another indicator of halting or stable patterns is when the population size stabilises. Another issue is how can the spatial output patterns be converted into formats that can be stored on classical computer systems? Again, a mechanism we have used is simple image analysis. Techniques such as automatic thresholding, image binarisation, hole counting and skeletonisation have been useful to encode the spatial pattern of network paths and nutrient locations into edges and nodes, respectively. How efficient and precise are these methods when compared to classical algorithms? This is still an open question as much of the research has focused on the novelty and control aspects but it is worth noting that these difficulties apply equally to the multi-agent material computation approach as they do to other unconventional computing substrates.

In concentrating on spatially implemented unconventional computation, this paper does not address the emergence of oscillatory phenomena in the multi-agent model. It bears mentioning, however, that a simple change in particle movement was sufficient to reproduce the emergence and dynamical transitions of oscillation patterns seen in *Physarum* [106], [107]. Furthermore, these oscillations could be harnessed to generate travelling waves in fixed populations, enabling collective transport [108] and amoeboid movement [109]. This has significant potential for the field of soft-robotics since, like *Physarum*, the model and its simple component parts effortlessly exploits the notion of distributed motor and control functions [110].

There may be potential in using the multi-agent approach in other areas apart from unconventional computing. From a more general perspective the multi-agent, virtual material approach makes a contribution to the study of pattern formation where previously studies have been dominated by numerical models or cellular automata. We have demonstrated that a wide range of reaction-diffusion patterning is possible using a multi-agent approach. The model yields a rich variety of patterning which does not rely on two-stage reactions in order to generate certain pattern types, as necessary in other approaches [111, 112]. It is notable that particles with identical shapes but opposite behaviour of those considered in this paper (where particles are *repelled* by chemoattractant) also show very complex pattern formation and evolution (see, [63] and [62] for further details). There is also potential for complex patterning by agents with completely different shapes (e.g. particles with fewer, or more, sensors, arranged in different architectures), suggesting that the simple agent-based method of dynamical pattern formation may generate interesting behaviours, or applications, to be explored in further research.

**Acknowledgements**

This research was supported by the EU research project "Physarum Chip: Growing Computers from Slime Mould" (FP7 ICT Ref 316366).